\documentstyle[preprint,aps]{revtex}

\begin{document}
\draft
\input{epsf}
\preprint{HEP/123-qed}
\title{Problems connected with electrons trajectory in separated atom}
\author{Evgueni V. Kovarski}
\address {
e-mail: ekovars@netscape.net}
\date{November 02, 2000}
\maketitle
\begin{abstract}
The time-dependent electromagnetic
field can results both pair waves and
pair particles.
It can be for mathematical relations between two functions
with identical argument and difference of phases equal
to $\pi$. Two examples both the opportunity of phase synchronism
for different frequencies of at atom -field interaction with known abnormal dispersion of refraction and
the possible trajectories for pair particles are considered.
\end{abstract}

\pacs{0155+b}

\narrowtext

\section*{Introduction}

At atom-field interaction frequency-time splitting of a spectral line
of spontaneous radiation can occurs due to given
the dressed atom theory at high intensity of external field
or for explained splitting each of two energy
atomics levels to some sublevels at low intensity of external field as a result
of quantum interference effect, as it was
supposed and demonstrated with help of 3D pictures titled as
Probability-Time-Frequency (PTF) manifolds ~\cite{E2}.
On the other hand for such observable characteristics as
the spontaneous radiation is possible to use the transformation
from trigonometrical functions of probability of a finding a
particle at a concrete energy level
inside atom to the exponential characteristics in space outside of atom.
The rate of emission for the quantum transition depend on the
derivative of probability $\it{P}(\tau)$
\begin{equation}
\it{R} =
 \int\limits^\infty_{0}\gamma\cdot
 \left[\it{exp}\left(-\gamma\it{t}\right)\right]\cdot
 \left(\dot{P}\right)\cdot\it{d}\it{t} \\
\end{equation}
The well known formula  for the
transition probability $\it{P}(\tau)$ given in single frequency
excitation of the two level atom is:
\begin{equation}
P_{1}=\left[\frac{ 4\Omega^2}{ 4\Omega^2+ \Delta\omega^2}\right]
\sin^2 \left(\frac{\tau}{2}\sqrt{4\Omega^2 +\Delta\omega^2}\right) \\
\end{equation}\\
where $\Omega$ is the Rabi frequency and
$\Delta\omega =\omega_0 - \omega_1,$ are the laser frequency detuning values.
Then the procedure of finding the rate of the power absorption
leads to the correspondent profile:
\begin{equation}
 R_{1} =\frac{ 2 \Omega^{2}\gamma}{\Delta\omega^{2}+
 4\Omega^{2}+\gamma^{2}} \\
\end{equation}
At a denominator of the formula there are three members,
one of which can be incorporated with other of two stayed members.
In this case this formula can be compared with a standard structure of
Lorentz profile, at which denominator always there are only two members.
Therefore communication of two spaces both the atomic and the external are
carried out except the amplitude in unique parameter known as a
full width on half of maximum (FWHM) of spectral line measured in a
scale of frequencies or time.
This FWHM for a Lorentz profile is designated
as $(2\gamma)$. The connection between this size of FWHM and a damping
constant of oscillator same designated, but describing losses of energy
of a particle at interaction with other particles in environment is especially
brightly shown for wide of FWHM spectral lines due to collisions with other atoms.
It is much more difficult to explain such
connection between FWHM and $(\it{2}\gamma)$ for isolated oscillator
or for radiation from stopped atom by laser cooling.
Despite of it isolated oscillator has spontaneous radiation and
therefore is characterized by so-called natural width of a spectral line
connected with the same $(\it{2}\gamma)$ as the damping
constant from classical equation of oscillator.
It is possible to tell that it is one of difficult examples when the
classical performances need to be combined with the quantum, if
to mean that the energy level can be split as in the dressed theory.
If such splitting is not connected to a magnetic field, therefore
 does not exist rules of selection for optical transitions and then
there is a question about that occurs with a particle on line
interacted with EM field if it is with identical
probability at different energy statuses of sub levels.
The time of spontaneous emission $\it{t_{S}}$ from upper energy level
is well known:
\begin{equation}
t_{S} = \frac { 3\pi\hbar\epsilon_{0} c_{0}^3 }{\omega_{0}^3d_{21}^2}    \\
\end{equation}
The probability that an electron  with life,
having unknown time $(\it{t})$, will leave the upper energy level, and its
spontaneous radiation will fade with known constant $(\gamma^{-1}_{\it{S}})$
is defined by function
\begin{equation}
\it{G}_{1}=\gamma_{S}\cdot\it{exp}\left(-\gamma_{S}\cdot\it{t}\right)
\end{equation}
The probability that the upper level
will become empty with a known damping constant $(\gamma)$ during the
unknown time of spontaneous radiation $(\it{t}_{S})$ is:
\begin{equation}
\it{G}_{2}=\gamma\cdot\it{exp}\left(-\gamma\cdot\it{t}_{S}\right)
\end{equation}
At definition of complete probability of time of life
it is necessary to take into account both probabilities ~\cite{E2}
\begin{equation}
\it{G}=\it{G}_{1}\cdot\it{G}_{2}=\frac{1}{\it{t}\cdot\it{t}_{S}}\cdot
\it{exp}\left[-\frac{\it{t}^{2}+\it{t}^{2}_{S}}{\it{t}\cdot\it{t}_{S}}\right]
\end{equation}
The lifetime $\it{t}$  going from function $\it{G}_{2}$ is oversized the
lifetime $\it{t}_{L} = 2 \it{t}_{s}$  from the first distribution $\it{G}_{1}$
due to the contribution of no radiative decay that is important for a separated
atom.

Other classical example is under consideration of the phase and group
speed of the waves
interaction with a separated atom, because for only one wave of an external EM field,
as against a single pulse mode or mode of several EM waves, the group speed
of the single wave exist only as a mathematics value
and therefore does not influence on phase speed, if of course is
to consider reception of a single external wave from single external oscillator.
It is well known that the dispersion of an index of refraction
$n(\omega)$ named as the abnormal refraction is in a vicinity of
resonant frequency $(\omega_{0})$ at
interaction of an electromagnetic wave (EM) with two levels atoms
$\it{E}=\it{E}_{0}\cdot\it{cos}(\omega\cdot\it{t})$ ,
where the wave number is $\it{k}=\omega\cdot\it{n}/\it{c}_{0}$,
$(\it{E}_{0})$ is the amplitude
of an external EM wave, $(\it{c}_{0})$ is the speed of the light in vacuum.
Phase $(\it{v}_{f})$ and group $(\it{v}_{g})$ speeds of a wave involved
for an explanation of this effect are defined by the known relations
due to the wave number can be spread out in a number on
degrees of frequency is in a vicinity of resonant frequency $(\omega_{0})$:
\begin{equation}
\it{v}_{\it{f}}=\frac{\it{c}_{0}}{\it{n}(\omega)}=\frac{\omega}{\it{k}(\omega)}
\end{equation}
\begin{equation}
\it{v}_{g}=\frac{1}{\it{dk/d}\omega}=\frac{\it{c}_{0}}
{\it{n}+(\omega-\omega_{0})\cdot(\it{dn/d}\omega)}
\end{equation}
On the other hand such single wave can be submitted from external space as a set
waves and then the physical sense for application of group speed is kept for
single wave. However it is possible to assume and return situation,
when single external wave cooperates with set of particles having for example a
parallel trajectories.

\section{Whether there is a problem of an electrons trajectory  in atom?}
Last example can be served by application of mutual orientation
of a trajectory of a particle and field. Because the phase speed
 $(\it{v}_{\it{f}})$ depends on $(\omega-\omega_{0})$,
therefore, in particular, for two separated EM waves with frequencies detuned
symmetrical respect to the central frequency of quantum transition in atom,
named as sidebands modes of EM field, as known, that is possible to
supply different phase speeds of EM waves in same environment.
Is possible spatially to divide these waves and to direct them
towards each other. So, the atom placed in such field, will test
action of different forces $\it{f}_{1,2}=\it{dp}/\it{dt}$,
where $\it{p}=m\cdot\it{dx}/\it{dt}$, where $(\it{m})$ is the mass
of the particle,$(\it{x})$ is the displacement of the atom along the EM field.
When an atom is under perturbation of two symmetrical and
coherent laser frequencies ~\cite{E2}, therefore the frequency detuning conditions for such
bichromate laser waves are:
\begin{equation}
 E(\tau)= E_{0} \left[\cos (\omega_{1}\tau) + \cos(\omega_{2}\tau)\right] \\
\end{equation}
\begin{equation}
\Delta\omega_2 = \omega_2-\omega_0\\
\end{equation}
\begin{equation}
\Delta\omega_1 = \omega_0-\omega_1 \\
\end{equation}
For the special perturbation of the upper energy level there
are two symmetrical frequencies and we can use the relation:
\begin{equation}
\Delta\omega_{2}=\Delta\omega_{1} =\Delta\omega \\
\end{equation}
The transition probability is:
\begin{equation}
\ P_{2}=\left[\sin\left(\Omega\cdot\frac{\sin\Delta\omega\tau}{\Delta\omega}\right)\right]^{2} \\
\end{equation}
  With Bessel functions the probability $\it{ P}_{2}$  can be written:
\begin{equation}
\ P_{2}=\left[2 J_1(\rho)\sin\Delta\omega\tau+2 J_3(\rho)\sin 3\Delta\omega\tau+
 2 J_5(\rho)\sin 5\Delta\omega\tau + ..\right]^2   \\
\end{equation}
\begin{equation}
\rho=\frac{\Omega}{\Delta\omega}
\end{equation}
 The rate of radiation $ \it{R}_{2}$
can be obtained by the same way as above for a single frequency excitation.
After some algebra with Bessel functions and by use the known relation
\begin{equation}
\it{sin}\left[\it{sin}\Theta\right]=\it{2}\cdot\sum^{\infty}_{\it{k=0}}
\it{J}_{\it{2k+1}}(\it{z})\cdot\it{sin}\left[\it{(2k+1)}\cdot\Theta\right]
\end{equation}
there are two  presentations of the rate $ \it{R}_{2}$ :
\begin{equation}
   R_{2}=\frac{\gamma}{2}(\Delta\omega)^{2}\sum\left[
    2 J_{ 2 k + 2}( 2\rho)\frac{ \left( 2 n + 2\right)^{2}}
   {\left( 2 n + 2\right)^{2} \Delta\omega^{2}+ \gamma^{2}}\right]  \\
\end{equation}
\begin{equation}
   R_{2}=\frac{\gamma}{2}(\Delta\omega)^{2}\sum\left[
    2 J_{ 2 k + 2}( 2\rho)\frac{1}
   {\Delta\omega^{2}+ \left(\frac{\gamma}{ 2 n + 2}\right)^{2}}\right]  \\
\end{equation}
Here numbers $(\it{k}\not=0)$ worth at Bessel functions $(\it{J}_{\it{2k}})$
coincide with numbers $(\it{2n+2=2k})$, where $(\it{n=0,1,2,..})$
There are series with Lorentzian profiles. If the condition for
detuning is: $\Delta\omega\gg\gamma$, then the rate of the power absorption
$ \it{R}_{2}$ is simply :
\begin{equation}
 R_{2} = \frac{\Gamma}{2}\left[1 - J_{0}\left( 2\rho\right)\right],  \\
\end{equation}

Thus the trajectory of an atom plays the important role, because the atom
can be stopped by sidebands method that results in known now results on
cooling atoms. Contrary an atom can be accelerated.

Let us note, that is interesting to consider the problem of the movement of
an electron within atom with the trajectory along and against a
direction of a wave vector of the EM field,
because with the rotation inside atom the electron can
additional braking or acceleration. So both the trajectory of electrons
within atom especially in a case of quantum transition and
spatial orientation of a field inside atom are important too.
In this work it is offered to discuss only some of these problems due to
this question is connected with the important presence of
a damping coefficient $(\gamma)$ in equation of the oscillator
and because such modes of movement should be shown in radiation
of a driven charge.

The quantum mechanics does not recognize exact definition of coordinates
or trajectory of movement. If the force in classical physics is defined as
$\it{f}=\it{dp}/\it{dt}$, in the quantum mechanics it is
$(\it{F}=-\partial\it{U}/\partial\it{t})$, where $(\it{U})$ represents
known potential. At the same time potential depends on coordinate $(x)$,
which by virtue of known uncertainty relation in the quantum mechanics
cannot be defined. However, a wave function $(\psi)$  in the quantum mechanics
describes wave with not observable characteristics as
the "wave frequency" $(\it{H}/\hbar)$ and is presented through the
operators of coordinate $(\it{X})$ and moment $(\Lambda)$,
because the hamiltonian $\it{2H}=\it{X}^{2}+\Lambda^{2}$:
\begin{equation}
\psi(\it{x,t})=\left[\it{exp}\left(-\imath\frac{\it{H}}{\hbar}\it{t}\right)\right]
\cdot\psi(\it{x,0})
\end{equation}
Application of the operators is
connected with two linear spaces, one of which is the space
of laboratory where classical physics is work and coordinate
 is measured as $\it{x}$:
\begin{equation}
\it{X}\psi=\it{x}\psi
\end{equation}
\begin{equation}
\Lambda\psi=-\imath\cdot\hbar\frac{\partial\psi}{\partial\it{x}}
\end{equation}
The information about space of an atom can be received only in space of
classical physics named in this work as the space of laboratory
where for the known classical model of the oscillator the exam of the
differential equation gives infinite on size value of displacement $(\it{x})$
of the oscillator for both conditions the absence of the damping constant
$(\gamma)$ and the exact resonance condition $(\omega=\omega_{0})$:
\begin{equation}
\ddot{x}+\it{2}\gamma\dot{x}+\omega^{2}_{0}\it{x}=\frac{\it{e}}{\it{2m}}
\cdot\it{E}_{0}\cdot\left[\it{exp}(\imath\omega\it{t})+
\it{exp}(-\imath\omega\it{t})\right]
\end{equation}
\begin{equation}
\it{x}=\left[
\frac{\it{e}}{\it{2m}}\cdot
\frac{\it{E}^{2}_{0}}
{\omega^{2}_{0}-\omega^{2}+ 2\imath\gamma\omega}\right]+\it{c.c.}
\end{equation}
Therefore it is obvious, that it is simultaneously impossible to simulate
both conditions $(\omega=\omega_{0})$ and $(\gamma=0)$.
Despite of it, it is possible to explain increase of displacement up to
infinite size by display of the appeared acceleration.
Really, the well known  equation of a clasical oscillator can be considered
 as a threshold condition at balance of  several acceleration mechanisms
for one connected electron. When the frequency of a field comes nearer to
resonant frequency, then in this vicinity of frequencies or in the
appropriate vicinity of time in system necessarily there should be an the
damping, differently irrespective of energy of an electromagnetic field
oscillations will simply be stopped when the electron becomes free,
because increase of displacement up to infinite size.

Therefore there is a vicinity of time, when there are processes with
several types of the damping mechanisms braking the electron.
From known observable processes of these  mechanisms, connected to presence,
first of all, it is necessary to allocate the absorption and both
spontaneous and compelled radiation.
The important problem for one atom is the understanding of a nature of
the spontaneous radiation damping constant.
This problem concerns to a question on so-called optical friction and
is general for the classical and quantum approach, as is connected
that the damping is present at processes, which are not for separately
taken atom, because there is no interaction with  environment.

In the quantum mechanics the damping $(\gamma)$ is entered in the
equations provided that  the electron already is at the exited level,
and the field at this moment has no energy or is switched simply off
like a case for a short pulse of a EM field.
The presence of the damping is entered in the assumption,
that probability of a finding  atom in the exited status is  .
In case of unlimited quantity of oscillators, the system of the
equations results for complete probability of spontaneous
transition from the exited status ~\cite{E1}. Thus the damping has the
imaginary part and gives the amendment on displacement of own frequency,
neglecting with which, however, it is possible to receive,
that is defined by summation on all radiated quantum.

The important note is about the time as the complex number, which
has not only size opposite to the frequency of an external field,
but also size, opposite to the complex damping coefficient.

The connection between a spectral contour for probability and its
FWHM is determined as well as by Lorentz function with inknown
value of the amplitude:
\begin{equation}
\it{L}=\it{const}\cdot\frac{\gamma}{\Delta\omega^{2}+\gamma^{2}/4}
\end{equation}
Let's note, that the size $(\Delta\omega=\vert\omega-\omega_{0}\vert)$
is variable and depends on adjustment of frequency of an external field
 $(\omega)$ in relation to resonant frequency of quantum transition
$(\omega_{0})$ in atom between two energy levels.
Let's consider the norm one of the two typical functions,
which make refraction index:
\begin{equation}
\it{k}_{\it{A}}= \it{const}\cdot\frac{\it{b}}{\it{a}^{2}+\it{b}^{2}}
\end{equation}
\begin{equation}
\it{k}_{\it{B}}= \it{const}\cdot\frac{\it{a}}{\it{a}^{2}+\it{b}^{2}}
\end{equation}
It is possible to fix in them one of two variable $(\it{a})$  or $(\it{b})$.
The norm can be used in three cases
\begin{equation}
\it{const}\cdot\int{\frac{\it{da}}{\it{a}^{2}+\it{b}^{2}}}=
\it{const}\cdot\left[\frac{1}{\it{b}}\it{arctan}\frac{\it{a}}{\it{b}}+\it{C}\right]
\end{equation}
\begin{equation}
\it{A}\int{\frac{\it{dy}}{\it{y}^{2}+1}}=\it{A}\cdot\pi=\it{1}
\end{equation}
\begin{equation}
\it{A}\cdot\pi=\int{\frac{\it{sin}^{2}(\it{A}\cdot\it{y})}{\it{y}^{2}}\cdot\it{dy}}
\end{equation}
Anyone here is taken variable $(y)$ and constants $(\it{A})$ and a free
constant $(\it{C})$ .
It is clear, that it is possible to solve this system of the equations
and to receive a number of the decisions, instead of one known $(\it{y})$.

Considering only one analytical function of complex variable $(\it{z})$
on a complex plane one can observe the power numbers $(\it{q})$ ~\cite{E3}:
\begin{equation}
\it{f(z)}=\frac{1}{\it{z}^{2}+1}
\end{equation}
In a circle $(\it{z}<\it{1})$, except for points $(\it{z}\not=\imath)$,
this function is an indefinitely decreasing geometrical progression
with the radius of convergence $(\it{r=1})$:
\begin{equation}
\it{f(z)}=\sum^{\infty}_{\it{q=0}}
\left(\it{-1}\right)^{\it{q}}\cdot\it{z}^{\it{2q}}
\end{equation}
In the other circle $\vert\it{z-1}\vert<\sqrt{\it{2}}$ ,
the radius of convergence $(\it{r=}\sqrt{2})$:
\begin{equation}
\it{f(z)}=\sum^{\infty}_{\it{0}}\left(\it{-1}\right)^{\it{q}}\cdot
\frac{\it{sin}(\it{q+1})\cdot\frac{\pi}{\it{4}}}{\it{2}^{\frac{\it{q+1}}{2}}}
\cdot\it{(z-1)}^{\it{q}}
\end{equation}
Therefore any quantum process observable in space of laboratory
and described by the formula passes through a number
$(\it{q}\not=\it{n}_{\it{0}})$, where $(\it{n}_{\it{0}})$
is the main quantum number of discrete energy levels. Consequently
the picture of energy levels distribution in the space of laboratory
should correspond to this picture.  It is possible to try to show
importance of distinction of performances about the main quantum number
in different spaces on the following example which is connected
with classical parameter of refraction.

The direct transition from consideration of one atom to consideration of
environment consisting many $(\it{N})$ atoms and back requires a physical
explanation of applicability for separated atom $(\it{N=1})$ not only
concept of the
damping constant, but also such concepts as a parameter of
refraction $(\it{n})$ or permeability $(\epsilon)$.
As is known, the polarization of  environment $(\it{P})$
is defined by the dipole moment $(\it{d=e}\cdot{\it{x}})$.
On the other hand, the polarization is defined by a susceptibility $(\alpha)$
and electrical component of an electromagnetic field $(\it{E})$:
\begin{equation}
\it{P}=\it{N}\cdot\alpha\cdot\it{E}=\it{N}\cdot\it{e}\cdot\it{x}=\it{N}\cdot\it{d}
\end{equation}
\begin{equation}
\epsilon=\it{1+4}\pi\cdot\alpha
\end{equation}
\begin{equation}
\it{n}=\sqrt{\epsilon}
\end{equation}
However permeability $(\epsilon)$ and parameter of refraction $(\it{n})$
can not be simply transferred from space of environment into the space of
separate atom.
 In this work is possible to assume,
that there are features of structure of atom,
which are shown that the connected electron in atom and the
free electron in space of laboratory are the same particle
which are taking place in different spaces.
Therefore they can have differences in such characteristics, as a charge
$(\it{e})$ and the mass $(\it{m})$,
because they have differences on dynamics of the movement driven by EM field.
The space of separate atom can have especial structure,
in which the mass and the charges of the connected electron interacted with a
EM field differ among themselves.
Thus connected electrons with a negative charge in different atoms are
considered taking place in different spaces, and free electrons
with a negative charge leaving of energy limits of different atoms
in space of laboratory are considered identical (standard electron).
The change of particles mass $(\it{m})$ during movement should result
in the certain conformity between complete energy of a particle and
energy of the appropriate quantum transition.
It is natural to assume, that the electromagnetic wave crosses set of
spaces with different phase speed in each of them.
Let's name as resonant spaces such spaces, which are connected among
themselves by any interaction to occurrence of a particle or information
electromagnetic field. As an information electromagnetic field we shall
name such electromagnetic field, which satisfies with the Maxwell
equations and bears in the characteristics the information on
investigated space.
    Let's assume, that the resonant spaces having, at least,
linear communication between one vector in one space with other vectors
in the other space cooperate only.
   Let's assume, that the birth of an electromagnetic field occurs on
border of resonant spaces.
   The communication of two linear spaces of laboratory and atom is
carried out by conformity between dependence of amplitude of a field
on time and frequency in one space and in the another space
with dependence of probability of finding electron at a energy level
from time and frequency.
   Factor of proportionality defines a time scale in each space, i.e.
the frequency of a EM field  $(\omega)$ and Rabi frequency $(\Omega)$:
\begin{equation}
\Omega=\frac{\it{d}_{12}\cdot\it{E}}{\hbar}
\end{equation}
where the dipole matrix element is $(\it{d}_{12}=\it{e}\cdot\it{x}_{12})$.
It is possible to assume, that the vectors $(\vec{\it{d}})$ and
$(\vec{\it{x}})$ characterize the space not as linearly dependent pair vectors,
and as linear - independent pair vectors, therefore is possible to write down:
\begin{equation}
\vec{\it{e}}=\it{e}_{1}\cdot\vec{\it{d}}+\it{e}_{2}\cdot\vec{\it{x}}
\end{equation}
Thus, the charge $(\it{e})$ of the connected electron is a vector
$(\vec{\it{e}})$. This vector has projections to the allocated directions
and, hence, brings in charging symmetry to space of atom.
Therefore, the application of the classical physical characteristics
such as $(\it{n})$, $(\epsilon)$ for separate atom is possible.

Assuming, that trajectory of an electron with speed $(\it{V})$ is the circle,
in which centre is nucleus, is necessary to return to
a hypothesis of Bohr and De Brogle. On one orbit with length $(\it{2}\pi\it{R})$,
where radius $(\it{R})$ is defined by:
\begin{equation}
\it{R}=\frac{\it{n}_{0}\cdot\hbar}{\it{m}\cdot\it{V}}\it{,}
\end{equation}
is possible to lay an integer of lengths of waves $(\lambda\cdot\it{n}_{0})$,
where $(\lambda)$ is characterizes a wave nature of a material particle
and $(\it{n}_{0})$ is the main quantum number:
\begin{equation}
\frac{\lambda}{\it{n}_{0}}=\frac{\it{h}}{\it{m}\cdot\it{V}}
\end{equation}
On the other hand  the same radius $(\it{R})$ of an orbit can be
determined from known relation:
\begin{equation}
\frac{1}{ 8\pi\epsilon_{0}}\cdot\frac{\it{e}^{2}}{\it{R}}=
\frac{\it{m}\cdot\it{V}^{2}}{ 2}
\end{equation}
One can to observe that the charge has the same vector nature that the velocity.
Let's copy the formula in view of a known ratio
$(\lambda\cdot\nu=\it{c}_{0})$, where the frequency is $(\it{2}\pi\nu=\omega)$,
the speed of light $(\it{c}_{0})$ is connected to phase speed of
electrons wave through $(\it{c}_{0}=\it{v}_{\it{f}}\cdot\it{n})$,
where $(\it{n})$ - is a parameter of refraction.
\begin{equation}
\it{n}=\frac{\hbar\omega}{\it{m}}\cdot\frac{1}{\it{v}_{\it{f}}\cdot\it{V}}\cdot\it{n}_{0}
\end{equation}
Thus, the parameter of refraction $(\it{n})$ of an atom should be
discrete size due to the main quantum number.

Let's return to
the decision of the classical equation of oscillator.
It is possible to notice what exactly from here follows,
that the permeability  $(\epsilon)$ is complex number.
For this reason the parameter of refraction $(\it{n}=\sqrt{\epsilon})$
is defined by a root of the second degree from complex number.
However parameter of refraction defines a phase speed of an electrons wave.
It is easy to show, that this phase speed is split on two waves
distinguished on a phase on number $(\pi)$.
This implies, that the main quantum number $(\it{n}_{0})$
also can result in processes with the same phase displacement.

In physics some cases are known, when the difference of phases is equal to
$(\pi)$.
First it is shown at reflection of a EM wave. Secondly it is possible to
search for such difference in phases at display of linear polarization of a wave.
In third, such difference exists between positive and negative parts of
periodic sine wave function.

\section{Possible pair processes for waves and particles}

Let's consider as an example probable reaction of an atom space
to action of an external field.
For this purpose we shall lead analogy between space of an
optical crystal and the two levels atom.  Let external field has given,
for example, ordinary linear polarization named  as a $\it{(o)}$ -wave,
which in turn, as is known, can be submitted consisting from both
the right and the left rotating vectors of polarization.
Such two vectors, opposite on rotation can be considered, as
two waves having the identical module and arguments distinguished
on number $\pi$.
On an entrance of a crystal such external $\it{(o)}$-field is usually represented
 as two identical $\it(o)$-waves having identical frequency $\omega$  and
two identical
amplitudes. Actually by such consideration the one wave goes about
two waves having the identical module and different arguments.
As a result of interaction, a unusual $\it{(e)}$-wave with the polarization
revolved on corner of 90 degrees is left the crystal.
As is known, there is such spatial direction inside a crystal, in which
the so-called  $\it{(oo-e)}$ interaction results in identical factors
of a parameter of refraction $\it{n}_{\it{0}}(\omega)=\it{n}_{\it{e}}(2\omega)$
 on the frequency $\omega$ for $\it{(o)}$- wave
 and on the frequency $2\omega$ for $\it{(e)}$ -wave
for known procedure of the second harmonic generation (SHG).
Both fields are external in relation to space of a crystal,
as their characteristics are measured in space of classical physics
named as a space of laboratory. Thus, the phase synchronism of speeds
$\it{v}_{\it{f}}=\it{c}_{0}/\it{n}$ is a reaction of interaction of two spaces.

In  gas environment of atoms the reaction of interaction
 both spaces the laboratory space and the space of atoms shown as
similar SHG of $2\omega$  should not
essentially differ from reaction by the space of a crystal.
Near to a line of absorption
always there is an opportunity of a choice of two frequencies with an
identical parameter of refraction due to the known abnormal dependence
of the refraction parameter.
In the specific case frequencies can be multiple, as SHG, but the square-law
dependence for low intensity external EM field of SHG intensity completely not
necessarily should be observed.
Thus it is necessary to distinguish, that at such low intencities the SHG
in atomic environment differs from known process of the two photon
absorption within an atom with the subsequent radiation.
Generally is possible to think that at high intencities the multi photons
generation of radiation differs from multi photons of absorption with
the subsequent radiation, because can occur at conditions when the determining
factor is the dependence $\it{n}(\omega)$ near to a line of resonant absorption.

External in relation to atom and periodically time-dependent
electromagnetic (EM) field $\it{E}= E_{o}\cos\omega\cdot\it{t}$
in space of laboratory is usually represented as
analytical continuation of function valid variable $(\omega\cdot\it{t})$
in complex area. In this work is paid attention to some features of such
transition in a complex plane and back from the point of view of
interpretation of physics of model connected to generation of an
electromagnetic field.

When on some given piece $\left[\it{a,b}\right]$,
 where $\it{a}=\omega_{1}\it{t}_{1}$ and  $\it{b}=\omega_{2}\it{t}_{2}$
of the valid real axis chosen along any direction of the space, that
for length $\left[\it{a,b}\right]$, included in complex area,
there is a unique function of the complex variable $\it{z=a}+\imath\cdot\it{b}$.
Accepting the same value that function
real variable the electromagnetic field can be submitted on this
piece by a converging sedate number in a known kind
\begin{equation}
\sum{\it{c_{n}}\left[\omega t-\omega_{0} t_{0}\right]^{\it{n}}}
\end{equation}
From the mathematical point of view on definition of complex numbers follows,
that one complex number can characterizes one physical value of pair real
numbers with the established order of following of one real number behind
another, where thus the only imaginary number $\it{z}=\imath\cdot\it{b}$
is equivalent to $\it{a}=0$.
In particular, the EM field in space of laboratory is submitted as
\begin{equation}
\frac{\it{E}_{0}}{2}\left[\it{exp}\left(\imath\cdot\omega\it{t}\right)+
\it{exp}\left(-\imath\cdot\omega\it{t}\right) \right]
\end{equation}
Also this EM field incorporates the sum of two functions complex variable
\begin{equation}
\zeta_{1,2}=\it{exp}\left(\it{z}_{1,2}\right)=\it{exp}\left(\it{a}\pm\imath
\cdot\it{b}\right)=\it{exp}\left(\pm\imath\cdot\omega\it{t}\right)
\end{equation}
It means, that two variable $\it{z}_{1,2}$ are correspond to a separately
taken wave where the module of each is identical,
but the arguments differ is familiar
\begin{equation}
\it{E}_{0}\vert\zeta\vert=\it{E}_{0}\it{exp}\left(\it{a}\right)=\it{E}_{0}
\end{equation}
\begin{equation}
\it{arg}\zeta_{1,2}=\pm\it{b}=\pm\omega\it{t}
\end{equation}
It is understandable, that interpretation of different marks of arguments
 of a field can serve performance about two vectors directed along the
allocated direction of a numerical axis in opposite directions and
distinguished on a corner equal $\pi$.

It is known, that the loss of energy of an external field communicates,
for example, with such effects, as effect of saturation in two-level atom
$(\it{N}_{1}=\it{N}_{2})$ or with other
known nonlinear effects occurring to a plenty connected  electrons
in different atoms at a large intensity of the external field.
At the same time, in separately taken atom connected electron
increases the energy only for the account $(\hbar\omega)$.
Therefore according to the quantum theory for excite separately
taken oscillator it is enough to have vaguely weak on intensity
external EM field.

Actually a threshold condition should exist as a ratio between
losses of the electron on radiation $\gamma_{\it{R}}$ and losses of
 a field on absorption $\gamma_{\it{A}}$.
At such understanding of processes the connected electron always radiates,
including at the moment of transition to the exited energy level.
For example, considering one-dimensional classical oscillator at
the included external field, according to the theory of Einstein,
it is necessary to exam processes of absorption, and also processes of
the compelled and spontaneous radiation.
These processes are characterized in
known coefficients $\it{A}_{\it{21}}$,$\it{B}_{\it{21}}$ and $\it{B}_{\it{12}}$ .
In this connection, it is possible to write down the balance equation
for various damping constants of classical oscillator, because
the damping constant of spontaneous radiation $\gamma_{\it{S}}$
is equal to $\it{A_{21}}$:
\begin{equation}
\it{A}_{21}+\frac{\it{E}_{\it{x}}^{2}}{\it{8}\pi}\it{B}_{21}=
\frac{\it{N}_{1}}{\it{N}_{2}}\frac{\it{E}_{\it{x}}^{2}}{\it{8}\pi}\it{B}_{12}
\end{equation}
\begin{equation}
\gamma_{\it{S}}+\gamma_{\it{B}}=\gamma_{\it{A}}
\end{equation}
The process of absorption is a resulting process and consequently the
return size of $\gamma_{\it{A}}$ corresponds to time of life
$\gamma_{\it{A}}=\it{t}^{-1}_{\it{L}}$.
Therefore the parameter $\gamma=\gamma_{\it{A}}$ should be present at the
equation of the oscillator. Therefore becomes obvious, that the time of life
should not be equaled of time spontaneous radiation $\gamma_{\it{S}}^{-1}$.

If there is an absorption of energy of EM field in space of atom, this
implies classical performance that the external field  should cross a
surface of atom under the laws by and large connected with the known laws
of Fresnel for environments.  Visible light absorbed by the majority of
known atoms, does not test of diffraction on such small on the sizes object
 as atom. However, for shorter lengths of waves or for atoms with the
large sizes occurring processes of reflection and (or) the absorption
would result in occurrence of a diffraction shadow from atom.
As the sizes of atom are small in comparison with length of a EM wave,
such sizes allow to make transition to a limiting case of disappearance
of objects in classical diffraction tasks, when the external field can be
submitted on a complex plane as a spiral, including in a limiting case
with a resulting vector along an imaginary axis.
Agreeing with application of a known principle of Optics it is possible
to assume, that the separate atom is a source of secondary waves.
Apparently, the secondary waves in interpretation of this principle
simply should mean the fact of an output from atom of pair waves with a
difference of phases distinguished on number $\pi$.

  The EM wave can be presented as a continuous analytical function
$\it{f(z)}$ of complex variable $\it{z}=\it{x}+\imath\cdot\it{y}$
with $\dot{\it{f}}\not=0$, through partial derivative of
functions $(\it{u})$ and $(\it{v})$:
\begin{equation}
\it{f}(\it{z})=\it{u}(\it{x,y})+\imath\cdot\it{v}(\it{x,y})
\end{equation}
The continuity of function is carried out during each period, when there
are limiting values $\it{lim} \left[\it{f(z)}\right]=\left[\it{f(z}_{0})\right]$
 in a vicinity $\it{z}\rightarrow\it{z}_{0}$, when the inverse function is
determined as
\begin{equation}
\it{z}=\phi\vert\it{f(z)}\vert
\end{equation}
\begin{equation}
\dot{\it{f}}=\frac{1}{\dot{\phi}\vert\it{f(z)}\vert}
\end{equation}
\begin{equation}
\left|\matrix{u_{x}& u_{y}\cr
               v_{x}& v_{y}\cr}\right|\not=0
\end{equation}
Therefore at return transition to the valid plane
 the function EM field as the function of complex variable
can be presented as two functions valid variable $(\it{u})$ and $(\it{v})$.

For presentation it is possible to consider possible interpretation of
 occurrence of a usual periodic wave with constant amplitude
$\it{E}= E_{o}\cdot\sin(\omega\it{t})$. Such recording for both
electrical components of a field means that at increase of size of product
of frequency at time there is an increase of a corner between these
components of a field. Here analogy to periodic movement on a circle for
 a particle having an electrical charge first of all is pertinent.
In this case it is possible to speak that between the center of a circle
where there is a positively charged nucleus and the particle with a negative
charge exists electromagnetic interaction resulting to such movement.
For a wave with a wave vector $\it{k}=\omega/\it{c}$, where $\it{c}$ is a speed
of light, such periodic decision corresponds to the wave equation
\begin{equation}
\frac{\partial^{2}\it{E}}{\partial\it{t}^{2}}+\it{c}^{2}\cdot\it{k}\cdot\it{E}=0
\end{equation}
If the particle is at fixed energy level, its trajectory can be connected to a
trajectory of other pair particles taking place on other energies
levels of the same atom. There is an example for a particle fixed
on a circle with a radius $\it{(a)}$ and driven together with a circle along
the allocated direction $\it{(x)}$ in space of an atom
in a plane $\it{(x,y)}$ with a parameter $\it{(T)}$ of the corner of
turn acts:
\begin{equation}
\it{x}=\it{a}\cdot\left[\it{T}-\it{sin(T)}\right]
\end{equation}
\begin{equation}
\it{y}=\it{a}\cdot\left[\it{1}-\it{cos(T)}\right]
\end{equation}
\begin{equation}
\it{x}=\it{Arccos}\frac{\it{a-y}}{a}\cdot\sqrt{\it{2ay}-\it{y}^{2}}
\end{equation}
\begin{equation}
\it{z}=\it{x}+\imath\cdot\it{y}=\it{a}\cdot\left[\it{T}+\imath\cdot\it{exp}\left(
-\imath\cdot\it{T}\right) \right]
\end{equation}
It is known, that such trajectory is a projection of movement on a spiral.
Radius of curvature monotonously changes in conformity with a sine function:
\begin{equation}
\rho=\left|\frac{\left[1+\left(\frac{\it{dy}}{\it{dx}}\right)
\right]^{\frac{3}{2}}}
{\frac{\it{d}}{\it{dx}}\left(\frac{\it{dy}}{\it{dx}}\right)}\right|
=\it{4a}\cdot\it{sin}\frac{T}{2}
\end{equation}
It is easy to notice connection between interpretation of a trajectory
for the simple and double frequency of a wave.
It is possible, that Radius of curvature $\rho$ of this trajectory for
pair fields is variable size and can corresponds to amplitude
$\it{(E)}$ of EM wave, the radius
of the circle $\it{(a)}$ can corresponds to $(\it{E}_{0})$, the corner $\it{T}$
can corresponds to $\omega\it{t}$. For pair particles it is possible to connect
these characteristics with the Rabi frequency.

In each point such curve radius of curvature is perpendicular trajectories
 and all together such radiuses form other curve, of what it is possible to be
convinced if to replace  $\it{T}=\tau -\pi$. This will be the same
trajectory, but with other coordinates:
\begin{equation}
\it{X+a}\cdot\pi=\it{a}\cdot[\tau-\it{sin}(\tau)]
\end{equation}
\begin{equation}
\it{Y+2a}=\it{a}\cdot[1-\it{cos}(\tau)]
\end{equation}
Therefore it is possible to assume, that there is one more field or particle
carrying out the same movement with the same frequency and time,
but taking place  on other circle with same radius.
The trajectories of these pair processes
will be moved with a phase difference on number $(\pi)$
and together they will correspond to one function with variable on a mark
by amplitude.
Such movement for a particle can be simulated as movement at a fixed energy
level with friction $(\gamma)$,
which is present at the equations of classical and quantum oscillators.
 Continuing this logic it is possible to enter other
parallel trajectories for charges of particles and it is possible to
investigate new EM processes.

In connection with the proposed model is possible to approve,
that if the order of following functions in investigated physical
process is established to within a constant difference of the phase,
thus is established possibility to observe a pair physical processes for
 waves and particles not only at space of the atom or the space
of laboratory, but for different spaces where the flat part of space in
general probably may be divided on inverse sub spaces with
in parallel current direct and return processes.

Probably here there should be balanced restrictions on imagination,
because it is possible too to assume, that the physical laws
containing mathematical functions from even degrees of the arguments
should correspond to pair processes and particles "koquark".

\end{document}